\documentclass[runningheads]{svmult}

\newcommand{\be}{\begin{equation}}
\newcommand{\ee}{\end{equation}}
\def\sm{M_\odot}
\def\etal{et al. }

\usepackage{makeidx}   
\usepackage{graphicx}  
\usepackage{subeqnar}  
\usepackage{multicol}  
\usepackage{physprbb}  
\makeindex             

\newcommand{\greeksym}[1]{{\usefont{U}{psy}{m}{n}#1}}
\newcommand{\umu}{\mbox{\greeksym{m}}}

\begin{document}
\title*{Primordial Black Holes as a Probe of Cosmology and High Energy Physics}
\toctitle{ Primordial Black Holes as a Probe of the Early Universe,
\protect\newline Gravitational Collapse, High Energy Physics and Quantum Gravity}
%
%
\titlerunning{Primordial Black Holes}
%
\author{B. J. Carr}
\authorrunning{B. J. Carr}
%
%
\institute{Astronomy Unit, Queen Mary, University of London,\\
Mile End Road, London E1 4NS, England}

\maketitle              

\begin{abstract}
Recent developments in the study of primordial black holes (PBHs) will be
reviewed, with particular emphasis on their formation and evaporation. PBHs
could provide a unique probe of the early Universe, gravitational collapse,
high energy physics and quantum gravity. Indeed their study may place
interesting constraints on the physics relevant to these areas even if they
never formed. 
\end{abstract}

\section{Introduction}

Hawking's discovery in 1974 that black holes emit thermal radiation due to quantum effects was surely one of the most important results in 20th century physics. This is because it unified three previously disparate areas of physics - quantum theory, general relativity and thermodynamics - and like all such unifying ideas it has led to profound insights. Although not strictly an application of quantum gravity theory, the theme of this meeting, it might be regarded as a  conceptual first step in that direction. Also there is a natural link in that the final stage of black hole evaporation, when the black hole is close to the Planck mass, can only be understood with a proper theory of quantum gravity.

In practice, only ``primordial black holes" which formed in the early Universe could be small enough for Hawking radiation to be important. Such a black hole will be referred to by the acronym ``PBH",  although this should not be confused with the acronym for ``Physikcentrum Bad Honnef", the institute hosting this meeting!  Interest in PBHs goes back nearly 35 years and some of the history of the subject will be reviewed in Section 2. As will be seen, interest was much intensified as a result of Hawking's discovery. Indeed, although it is still not definite that PBHs ever formed, it was only through thinking about them that Hawking was led to his remarkable insight. Thus the discovery illustrates that studying something may be useful even if it does not exist!

Of course, the subject is much more interesting if PBHs {\it did} form and their discovery would provide a unique probe of at least four areas of physics: the early Universe; gravitational collapse; high energy physics; and quantum gravity. 
The first topic is relevant because studying PBH formation and evaporation can impose important constraints on primordial inhomogeneities, cosmological phase transitions (including inflation) and varying-G models. These topics are covered in Sections 3, 4 and 5, respectively.  The second topic is discussed in Section 6 and relates to recent developments in the study of ``critical phenomena" and the issue of whether PBHs are viable dark matter candidates.
The third topic arises because PBH evaporations could contribute to cosmic rays, whose energy distribution would then give significant information about the high energy physics involved in the final explosive phase of black hole evaporation. This is covered in Section 7. The fourth topic arises because it has been suggested that quantum gravity effects could appear at TeV scale and this leads to the intriguing possibility that small black holes could be generated in accelerators experiments or cosmic ray events. As discussed in Section 8, this could have striking observational consequences. Although such black holes are not technically ``primordial", this possibility would have radical implications for PBHs themselves.

\section{Historical Overview}

It was realized many years ago that black holes with a wide range of masses could have formed in the early Universe as a result of the great compression associated with the Big Bang. A 
comparison of the cosmological density at a time $t$ after the Big Bang with the density associated with a black hole of mass $M$
shows that PBHs would have of order the particle horizon mass at their formation epoch:
\be
M_H(t) \approx {c^3 t\over G} \approx 10^{15}\left({t\over 10^{-
23} \ {\rm s}}\right) g.
\ee
PBHs could thus span an enormous mass range: those formed at the
Planck time ($10^{-43}$s) would have the Planck mass ($10^{-5}$g),
whereas those formed at 1~s would be as large as $10^5\sm$,
comparable to the mass of the holes thought to reside in galactic nuclei. 
By contrast, black holes forming at the present epoch could never be smaller than about $1\sm$.

Zeldovich \& Novikov \cite{zn} first derived eqn (1) but
they were really considering ``retarded cores" rather than black holes and Hawking \cite{h71} was the first person to realize that primordial density perturbations might lead to gravitational collapse on scales above the Planck mass. For a while the existence of PBHs seemed unlikely since Zeldovich \& Novikov \cite{zn} had pointed out that they might be expected to grow catastrophically. This is because a simple Newtonian argument  suggests that, in a radiation-dominated universe, black holes much smaller than the horizon cannot grow much at all, whereas those of size comparable to the horizon could continue to grow at the same rate as it throughout the radiation era. Since we have seen that a PBH must be of order the horizon size at formation, this suggests that all PBHs could grow to have a mass of order $10^{15}\sm$ (the horizon mass at the end of the radiation era). There are strong observational limits on how many such giant holes the Universe could contain, so the implication seemed to be that very few PBHs ever existed.

However, the Zeldovich-Novikov argument was questionable since it neglected the cosmological expansion and this would presumably hinder the black hole growth. Indeed myself and Hawking were able to disprove the notion that PBHs could grow at the same rate as the particle horizon by demonstrating that there is no spherically symmetric similarity solution which represents a black hole attached to an exact Friedmann model via a sound-wave \cite{ch}. Since a PBH must therefore soon become much smaller than the horizon, at which stage cosmological effects become unimportant, we concluded that PBHs cannot grow very much at all (cf. \cite{bh,lcf}).

The realization that small PBHs might exist after all prompted Hawking
to study their quantum properties. This led to his famous 
discovery \cite{h74}
that black holes radiate thermally with a temperature
\be
T = {\hbar c^3\over 8\pi GMk}
\approx 10^{-7} \left({M\over \sm}\right)^{-1}\ {\rm K},
   \ee
so they evaporate on a timescale 
\be
      \tau(M) \approx { G^2M^3 \over \hbar c^4}
\approx 10^{64} \left({M\over \sm}\right )^3~\mbox{y}.
   \ee
Only black holes smaller
than $10^{15}$g would have evaporated by the present epoch, so eqn (1) implies
that this effect could be important only for black holes which formed before $10^{-23}$s.

Despite the conceptual importance of this
result, it was bad news for PBH enthusiasts. For since PBHs
with a mass of $10^{15}$g
would be producing photons with energy of order 100~MeV at the present epoch, the observational limit on
the $\gamma$-ray background intensity at 100~MeV immediately 
implied
that their density could not exceed $10^{-8}$ times the
critical density \cite{ph}. Not only did this render 
PBHs
unlikely dark matter candidates, it also implied that there was 
little
chance of detecting black hole explosions at the present epoch \cite{pw}.
Nevertheless, it was realized that PBH
evaporations could still have interesting cosmological 
consequences. In particular, they might
generate the microwave background \cite{zs} or
modify the standard cosmological nucleosynthesis scenario \cite{npsz} 
or contribute to the cosmic baryon asymmetry \cite{b80}.
PBH evaporations might also account for
the annihilation-line radiation coming from the Galactic centre \cite{or} 
or the unexpectedly high
fraction of antiprotons in cosmic rays \cite{kir}. PBH explosions occurring
in an interstellar
magnetic field might also generate radio bursts \cite{r77}. 
Even if PBHs had none of
these consequences, studying such effects leads to strong upper limits on how
many of them could ever have formed and thereby constrains models of the early Universe.

Originally it was assumed that PBHs would form from initial inhomogeneities but in the 1980s attention switched to several new formation mechanisms. Most of the mechanisms were associated with various phase transitions that might be expected to occur in the early Universe and there was particular interest in whether PBHs could form from the quantum fluctuations associated with the many different types of inflationary scenarios. Indeed it soon became clear that there are many ways in PBHs serve as a probe of the early Universe and, even if they never formed, their non-existence gives interesting information \cite{c85}. In this sense, they are similar to other ``relicts" of the Big Bang, except that they derive from much earlier times.

In the 1990s work on the cosmological consequences of PBH evaporations was revitalized as a result of calculations by my PhD student Jane MacGibbon. 
She realized that the usual assumption that particles are emitted 
with a black-body spectrum as soon as the temperature of the hole 
exceeds their rest mass is too simplistic. If one adopts the conventional view
that all particles are composed of a small number of fundamental
point-like constituents (quarks and leptons), it would seem 
natural to assume that it is these fundamental particles rather than the
composite ones which are emitted directly once the temperature 
goes above the QCD confinement scale of 250~MeV. One can
therefore envisage a black hole as emitting relativistic quark and gluon jets
which subsequently fragment into leptons and hadrons \cite{m91,mw} and this modifies the cosmological constraints considerably \cite{mc}

Over the last decade PBHs have been assigned various other cosmological roles. Some people have speculated that PBH evaporation, rather than proceeding indefinitely,  could cease when the black hole gets down to the Planck mass \cite{bow,cpw}. In this case, one could end up with stable Planck mass relics, which would provide dark matter candidates \cite{bcl,cgl,m87}. Although most gamma-ray bursts are now known to be at cosmological distances, it has been proposed that some of the short period ones could be nearby exploding PBHs \cite{bel,csh}. Solar mass PBHs could form at the quark-hadron phase transition and, since some of these should today reside in our Galactic halo, these have been invoked to explain the microlensing of stars in the Magellanic Clouds \cite{inn,j97,y97}.

\section{PBHs as a probe of primordial inhomogeneities}

One of the most important reasons for studying PBHs is that it enables one to 
place limits on the spectrum of density fluctuations in the early Universe. This
is because, if the PBHs form
directly from density perturbations, the fraction of
regions undergoing collapse at any epoch is determined 
by
the root-mean-square amplitude $\epsilon$ of the fluctuations
entering the horizon at that epoch and the equation of state 
$p=\gamma \rho~(0< \gamma <1)$. One usually expects a radiation equation of
state ($\gamma =1/3$) in the early Universe. In order to 
collapse
against the pressure, an overdense region must be larger than the
Jeans length at maximum expansion and this is just 
$\sqrt{\gamma}$
times the horizon size. On the other hand, it cannot be larger than the horizon size, else it would
form a separate closed universe and not be part of our Universe \cite{ch}. 

This has two important implications. Firstly, PBHs forming at time $t$ should
have of order the horizon mass given by eqn (1).
Secondly, for a region destined to collapse to a PBH, one requires the fractional overdensity at the horizon
epoch $\delta$ to exceed $\gamma$. Providing the density 
fluctuations have a Gaussian distribution and are spherically
symmetric, one can infer that the fraction of regions of mass 
$M$ which collapse is \cite{c75}
   \be
\beta(M) \sim \epsilon(M) \exp\left[-{\gamma^2\over 
2\epsilon(M)^2}\right]
   \ee
where $\epsilon (M)$ is the value of $\epsilon$ when the 
horizon
mass is $M$. The PBHs can have an extended mass spectrum only 
if the fluctuations are scale-invariant (i.e. with $\epsilon$ 
independent of $M$). 
In this case, the PBH mass spectrum is given by \cite{c75}
\be
dn/dM = (\alpha -2) (M/M_*)^{-\alpha}M_*^{-2}\Omega_{PBH}\rho_{crit}
\ee
where $M_* \approx 10^{15}$g is the current lower cut-off in the mass spectrum due to evaporations, $\Omega_{PBH}$ is the total density of the PBHs in units of the critical density (which itself depends on $\beta$) and the exponent $\alpha$ is  determined by the equation of state:
\be
\alpha = \left({1+3\gamma \over 1+\gamma}\right) + 1 .
\ee
$\alpha=5/2$ if one has a radiation equation of state ($\gamma$=1/3), as expected. This means that the density of PBHs larger than $M$ falls off as $M^{-1/2}$, so most of the PBH density is contained in the smallest ones. 

Many scenarios for the cosmological density fluctuations predict that $\epsilon$ is at least approximately scale-invariant but the sensitive dependence of $\beta$ on $\epsilon$ means that even tiny deviations from scale-invariance can be important.  If $\epsilon (M)$ decreases with increasing $M$, then the spectrum falls off exponentially and most of the PBH density is contained in the smallest ones. If $\epsilon (M)$ increases with increasing $M$, the spectrum rises exponentially and - if PBHs were to form at all - they could only do so at large scales. However, the microwave background anisotropies would then be larger than observed, so this possibilty can be rejected.

The current density parameter $\Omega_{PBH}$ associated
with PBHs which form at a redshift $z$ or time $t$ is related to
$\beta$ by \cite{c75}
   \be
\Omega_{\rm PBH} = \beta\Omega_R(1+z) \approx 10^6 
\beta\left({t\over s}\right)^{-1/2} \approx 10^{18}\beta\left({M\over 10^{15}g}\right)^{-1/2}
   \ee
where $\Omega_R \approx 10^{-4}$ is the density parameter of the microwave
background and we have used eqn (1). The $(1+z)$ factor arises 
because the radiation density scales as $(1+z)^4$, whereas the PBH
density scales as $(1+z)^3$. Any limit on $\Omega_{PBH}$ therefore places a constraint on $\beta(M)$ and 
the constraints are summarized in Fig.1, which is taken from Carr et al. \cite{cgl}.
The constraint for non-evaporating mass ranges above
$10^{15}$g comes from requiring
$\Omega_{PBH}<1$ but stronger constraints are associated with PBHs
smaller than this since they would have evaporated by now \cite{c76}. 
The strongest one is the $\gamma$-ray limit associated with the $10^{15}$g PBHs evaporating
at the present epoch \cite{ph}. Other ones are associated with the generation of entropy and modifications to the cosmological production of light 
elements \cite{npsz}. The
constraints below $10^{6}$g are based on the (uncertain) assumption that evaporating PBHs leave stable Planck
mass relics, in which case these  relics are required to have less than the critical density \cite{bcl,cgl,m87}.

The constraints on 
$\beta(M)$ can be converted into constraints on 
$\epsilon(M)$ using eqn (4) and these are shown in Fig.2. Also shown here are the (non-PBH) constraints
associated with the spectral distortions in the cosmic microwave background induced by the dissipation of 
intermediate scale density perturbations and the COBE quadrupole measurement. This shows that one needs the
fluctuation amplitude to decrease with increasing scale in order to produce PBHs and the lines corresponding
to various slopes in the $\epsilon (M)$ relationship are also shown in Fig.2.
 
\begin{figure}
\centerline{
\includegraphics[width=10cm]{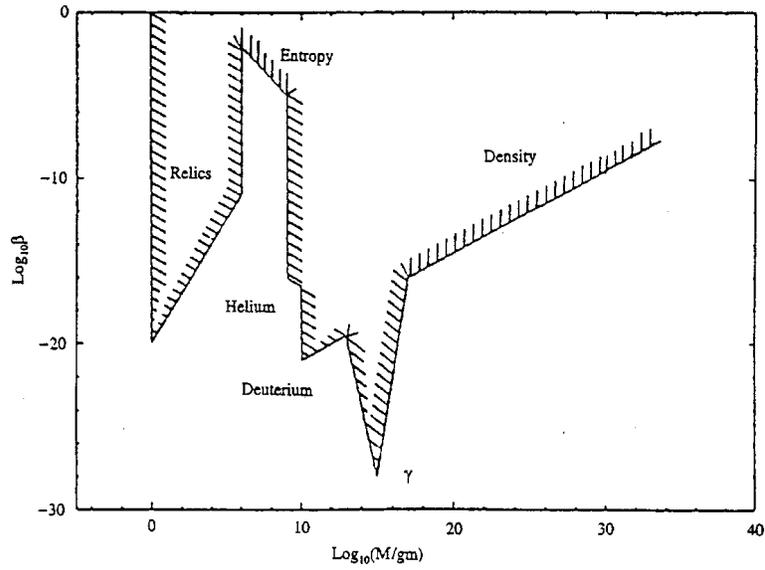}}
\caption{Constraints on $\beta(M)$}
\end{figure}

\begin{figure}
\centerline{
\includegraphics[width=10cm]{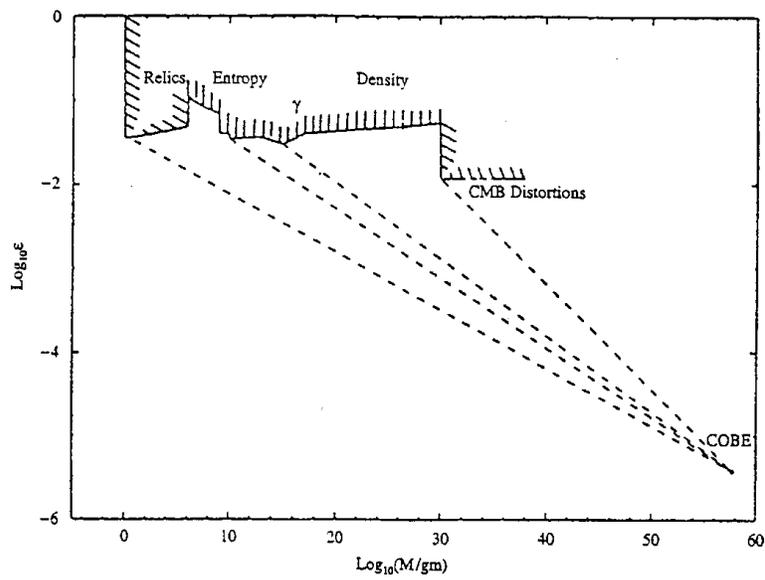}}
\caption{Constraints on $\epsilon(M)$}
\end{figure}

\section{PBHs as probe of cosmological phase transitions}

Many phase transitions could occur in the early Universe which lead to PBH formation. Some of these mechanisms still require pre-existing density fluctuations but in others the PBHs form spontaneously even if the Universe starts off perfectly smooth. In the latter case, $\beta(M)$  depends not on $\epsilon(M)$ but
on some other cosmological parameter.

\subsection{Soft equation of state}
Some phase transitions can lead to the equation of state becoming soft ($\gamma<<1$) for a while. For example, the pressure may be reduced if the Universe's mass is ever channelled into particles which are massive enough to be non-relativistic. In such cases, the effect of pressure in stopping collapse is unimportant and the probability of PBH formation just depends upon the fraction of regions which are sufficiently spherical to undergo collapse; this can be shown to be \cite{kp}
\be
\beta = 0.02 \epsilon^{13/2}.
\ee
The value of $\beta$ is now much less sensitive to $\epsilon$ than indicated by eqn (4) and most of the PBHs will be smaller than the horizon mass at formation by a factor $\epsilon^{3/2}$. For a given spectrum of primordial fluctuations, this means that there may just be a narrow mass range - associated with the period of the soft equation of state - in which the PBHs form. In particular, this could happen at the quark-hadron phase transition since the pressure may then drop for a while \cite{j97}.

\subsection{Collapse of cosmic loops}
 
In the cosmic string scenario, one expects some strings to self-intersect and form cosmic loops. A typical loop will be larger than its Schwarzschild radius by the inverse of the factor $G\mu$, where $\mu$ is the mass per unit length. If strings play a role in generating large-scale structure, $G\mu$  must be of order $10^{-6}$. Hawking \cite{h89} showed that there is always a small probability that a cosmic loop will get into a configuration in which every dimension lies within its Schwarzschild radius and he estimated this to be 
\be
\beta \sim (G\mu)^{-1}(G\mu x)^{2x-2}
\ee
where $x$ is the ratio of the loop length to the correlation scale. If one takes $x$ to be 3, $\Omega_{PBH}>1$ for $G\mu >10^{-7}$, so he argued that one overproduces PBHs in the favoured string scenario. Polnarev \& Zemboricz \cite{pz} obtained a similar result. However, $\Omega_{PBH}$  is very sensitive to $x$ and a slight reduction could still give an interesting value \cite{cc,gs,mbw}. Note that spectrum (5) still applies since the holes are forming with equal probability at every epoch.

\subsection{Bubble collisions}  

Bubbles of broken symmetry might arise at any spontaneously broken symmetry epoch and various people, including Hawking, suggested that PBHs could form as a result of bubble collisions \cite{cs,hms,ls}. However, this happens only if the bubble formation rate per Hubble volume is finely tuned: if it is much larger than the Hubble rate, the entire Universe undergoes the phase transition immediately and there is not time to form black holes; if it is much less than the Hubble rate, the bubbles are very rare and never collide. The holes should have a mass of order the horizon mass at the phase transition, so PBHs forming at the GUT epoch would have a mass of $10^3$g, those forming at the electroweak unification epoch would have a mass of $10^{28}$g, and those forming at the QCD (quark-hadron) phase transition would have mass of around $1\sm$. Only a phase transition before $10^{-23}$s would be relevant in the context of evaporating PBHs.

\subsection{Inflation}

Inflation has two important consequences for PBHs. On the one hand, any PBHs formed before the end of inflation will be diluted to a negligible density. Inflation thus imposes a lower limit on the PBH mass spectrum:
\be
M > M_{min} = M_{Pl}(T_{RH}/T_{Pl})^{-2}			 		
\ee
where $T_{RH}$ is the reheat temperature and $T_{Pl}\approx 10^{19}$ GeV is the Planck temperature. The CMB quadrupole measurement implies $T_{RH}\approx 10^{16}$GeV, so 
$M_{min}$ certainly exceeds $1$g. On the other hand, inflation will itself generate fluctuations and these may suffice to produce PBHs after reheating. If the inflaton potential is $V(\phi)$, then the horizon-scale fluctuations for a mass-scale $M$ are
\be
\epsilon(M) \approx [V^{3/2}/(M_{Pl}^3 V')]_H
\ee							
where a prime denotes $d/d \phi$  and the right-hand-side is evaluated for the value of $\phi$ when the mass-scale $M$ falls within the horizon. 

In the standard chaotic inflationary scenario, one makes the ``slow-roll" and ``friction-dominated" asumptions:
\be
\xi \equiv (M_{Pl}V'/V)^2 << 1,  \quad    \eta \equiv M_{Pl}^2 V''/V << 1	.	
\ee
Usually the exponent $n$ characterizing the power spectrum of the fluctuations, 
$|\delta _k|^2 \approx k^n$, is very close to but slightly below 1:
\be
n = 1 + 4\xi - 2\eta \approx 1.							
\ee
Since $\epsilon$ scales as $M^{(1-n)/4}$, this means that the fluctuations are slightly increasing with scale. The normalization required to explain galaxy formation ($\epsilon \approx 10^{-5}$) would then preclude the formation of PBHs on a smaller scale. 
If PBH formation is to occur, one needs the fluctuations to decrease with increasing mass ($n>1$) and this is only possible if the scalar field is accelerating sufficiently fast:
\be
V''/V > (1/2) (V'/V)^2  .							
\ee
This condition is certainly satisfied in some scenarios \cite{cl} and, if it is, eqn (4) implies that the PBH density will be dominated by the ones forming immediately after reheating. Since each value of $n$ corresponds to a straight line in Fig.2, any particular value for the reheat time $t_1$ corresponds to an upper limit on $n$.
This limit is indicated in Fig.3, which is taken from Carr et al. \cite{cgl} apart from a correction pointed out by Green \& Liddle \cite{gl97}. Similar constraints have now been obtained by several other people \cite{bkp,klm}. The figure also shows how the constraint on $n$ is strengthened if the reheating at the end of inflation is sufficiently slow for there to be a dust-like phase \cite{glr}.
PBHs have now been used to place constraints on many other sorts of inflationary scenarios - supernatural \cite{rsg}, supersymmetric \cite{g99}, hybrid \cite{glw,kky}, oscillating \cite{tar}, preheating \cite{bt,ep,fk,gm} and running mass \cite{lgl} - as well as a scenarios in which the inflaton serves as the dark matter \cite{lmu}.

\begin{figure}
\centerline{
\includegraphics[width=10cm]{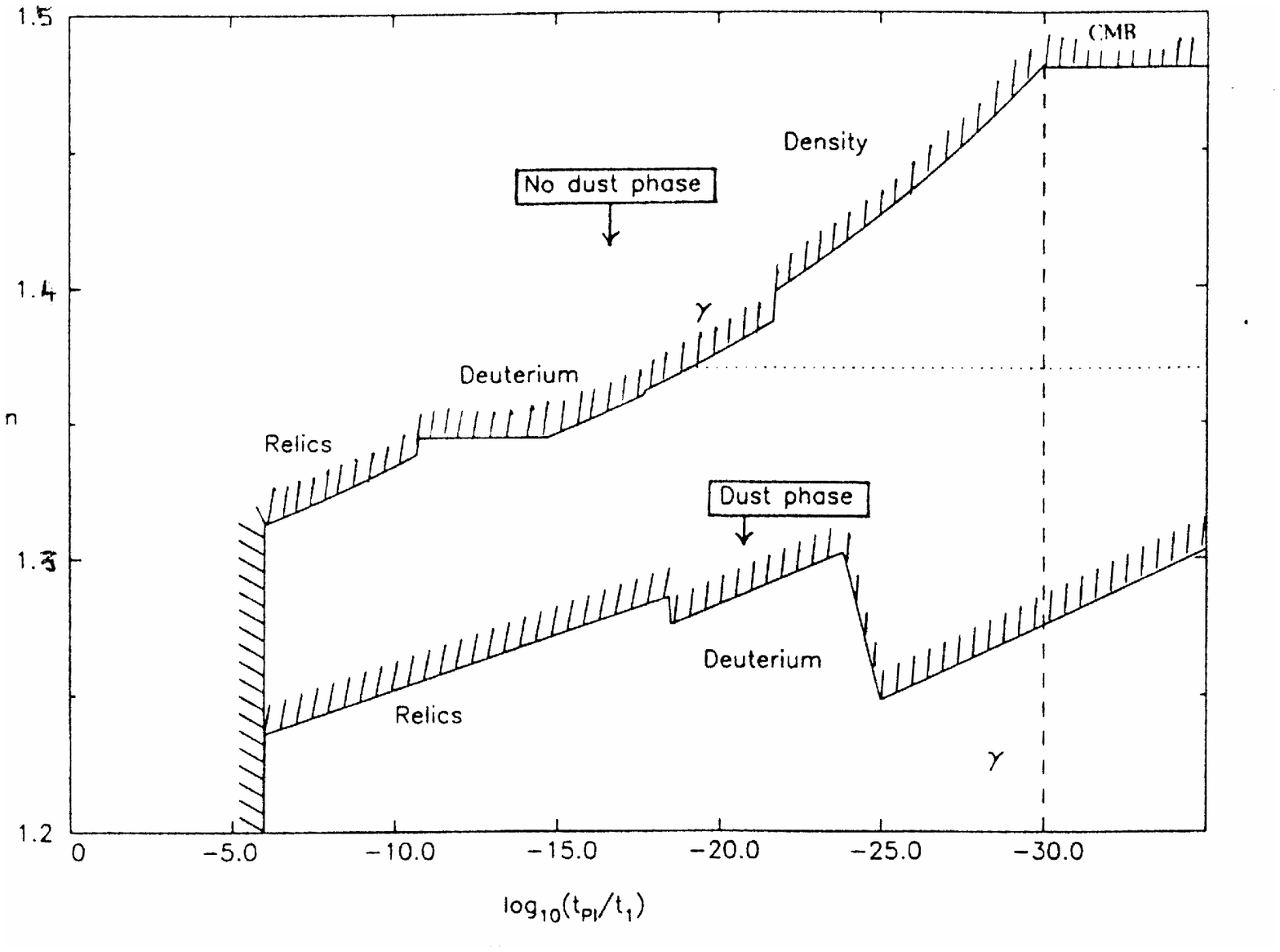}}
\caption{Constraints on spectral index $n$ in terms of reheat time $t_1$}
\end{figure}

Bullock \& Primack \cite{bp2} and Ivanov \cite{i98} have questioned whether the Gaussian assumption which underlies eqn (4) is valid in the context of inflation. So long as the fluctuations are small ($\delta \phi /\phi <<1$), as certainly applies on a galactic scale, this assumption is valid. However, for PBH formation one requires  $\delta \phi /\phi \sim 1$, and, in this case, the coupling of different Fourier modes destroys the Gaussianity. Their analysis suggests that $\beta(M)$ is much less than indicated by eqn (4) but it still depends very sensitively on $\epsilon$.

Not all inflationary scenarios predict that the spectral index should be constant. Hodges \& Blumenthal \cite{hb} pointed out that one can get any form for the fluctuations whatsoever by suitably choosing the form of $V(\phi)$. For example, eqn (11) suggests that one can get a spike in the spectrum by flattening the potential over some mass range (since the fluctuation diverges when $V'$ goes to 0). This idea was exploited by Ivanov et al. \cite{inn}, who fine-tuned the position of the spike so that it corresponds to the microlensing mass-scale.

\section{PBHs as a probe of a varying gravitational constant}

The PBH constraints would be severely modified if the value of the gravitational ``constant" $G$ was different at early times. The simplest varying-$G$ model is Brans-Dicke (BD) theory \cite{bd}, in which $G$ is associated with a scalar field $\phi$ and the deviations from general relativity are specified by  a parameter $\omega$. A variety of astrophysical tests currently require
$|\omega| > 500$, which implies that the deviations can only ever be small \cite{will}. However, there exist generalized scalar-tensor theories \cite{b68,nor,wag} in which $\omega$ is itself a function
of $\phi$ and these lead to a considerably broader range of variations in $G$. In  particular, it permits
$\omega$ to be small at early times (allowing noticeable variations of $G$ then) even if it is large
today. In the last decade interest in such theories has been revitalized as a result of early Universe studies.
Extended inflation explicitly requires a model in which $G$ varies \cite{ls} and,
in higher dimensional Kaluza-Klein-type cosmologies, the variation in the sizes of the extra dimensions also naturally
leads to this \cite{fr,kpw,mae}.

The behaviour of homogeneous cosmological models in BD theory is well understood \cite{bp1}. They are vacuum-dominated at early times but always tend towards the general relativistic solution during the
radiation-dominated era. This
means that the full radiation solution can be approximated by joining a
BD vacuum solution to a general relativistic radiation solution at some time which 
may be regarded as a free parameter of the theory. However, when the matter density becomes greater than the radiation
density at around $10^5$y,  
the equation of state becomes dustlike $(p = 0)$ and $G$ begins to vary again. 

The consequences of the cosmological variation of $G$ for PBH evaporation depend upon how the value of $G$
near the black hole evolves. Barrow \cite{b92} introduces two possibilities: in scenario A, $G$ everywhere maintains
the background cosmological value (so $\phi$ is homogeneous); in scenario B, it
preserves the value it had at the formation epoch near the black hole even though it 
evolves at large distances (so $\phi$
becomes inhomogeneous).
On the assumption that a PBH of mass $M$ has a temperature and mass-loss rate
\be
T \propto (GM)^{-1},\quad \dot{M} \propto (GM)^{-2},
\ee
with $G=G(t)$ in scenario A and $G=G(M)$ in scenario B, Barrow \& Carr \cite{bc} calculate how the evaporation constraints summarized in Fig.1 are modified for a wide range of varying-$G$ models.
The question of whether scenario A or scenario B is more plausible has been studied in several papers \cite{cg,gc,hcg,jac} but is still unresolved.

\section{PBHs as a probe of gravitational collapse}

The criterion for PBH formation given in Section 3 is rather simplistic and not based on a detailed calculation. The first numerical studies of PBH formation were carried out by Nadezhin et al. \cite{nnp}. These roughly confirmed the criterion $\delta > \gamma$ for PBH formation, although the PBHs could be be somewhat smaller than the horizon. In recent years several groups have carried out more detailed hydrodynamical calculations and these have refined the $\delta > \gamma$ criterion and hence the estimate for $\beta(M)$ given by eqn (4).
Niemeyer \& Jedamzik \cite{nj99} find that one needs $\delta >0.8$ rather than $\delta >0.3$ to ensure PBH 
formation and they also find that there is little accretion after PBH formation, as expected theoretically  \cite{ch}. Shibata \& Sasaki \cite{ss} reach similar conclusions.

A particularly interesting development has been the application of ``critical phenomena"
to PBH formation. Studies of the collapse of various types of spherically symmetric matter fields have shown that there
is  always a critical solution which separates those configurations which form a black hole from 
those which disperse to an asymptotically flat state. The configurations are described by some index $p$ and, as the
critical index $p_c$ is approached, the black hole mass is found to scale as $(p-p_c)^{\eta}$ for some exponent
$\eta$. This effect was first discovered for scalar fields \cite{chop} but subsequently
demonstrated for radiation \cite{ec} and then more general fluids with equation of state $p=\gamma \rho$
\cite{kha,mai}. 

In all these studies the spacetime was assumed to be asymptotically flat. However, Niemeyer \& Jedamzik \cite{nj98} have 
recently applied
the same idea to study black hole formation in asymptotically Friedmann models and have found similar results. 
For a variety of initial density perturbation profiles, they find that the
relationship between the PBH mass and the the horizon-scale density perturbation has the form
\be
M = K M_H(\delta - \delta_c)^{\gamma}
\ee
where $M_H$ is the horizon mass and the constants are in the range $0.34<\gamma<0.37$, $2.4<K<11.9$ and
$0.67<\delta_c <0.71$ for the various configurations. Since $M \rightarrow 0$ as $\delta \rightarrow \delta_c$, 
this suggests that PBHs may be much smaller than the particle horizon at formation and it also modifies the mass spectrum \cite{g00,gl99,klr,yok}. However, it is clear that a fluid
description must break down if they are too small and recent calculations by Hawke \& Stewart \cite{hs} show that black holes can only form on  scales down to $10^{-4}$ of the horizon mass.

There has also been interest recently in whether PBHs could have 
formed at the quark-hadron phase transition at $10^{-5}$s because of a temporary softening of the equation of state then.
Such PBHs would naturally have the sort of mass required to explain the MACHO microlensing results \cite{j97}. If the QCD phase 
transition is assumed to be of 1st order, then hydrodynamical calculations show that the value of
$\delta$ required for PBH formation is indeed reduced below the value which pertains in the radiation case \cite{jn99}. 
This means that 
PBH formation will be strongly enhanced at the QCD epoch, with the mass distribution being peaked around the horizon mass. One of the interesting
implications of this scenario is the possible existence of a halo population of {\it binary} black
holes \cite{nak}. With a full halo of such objects, there could then be $10^8$ binaries inside 50 kpc and some of these could be coalescing due to
gravitational radiation losses at the present epoch. If the associated gravitational waves were detected, it would provide a unique probe of the halo distribution \cite{itn}.

\section{PBHs as a probe of high energy physics}

We have seen that a black hole of mass $M$ will emit particles
like a black-body of temperature \cite{h75}
   \be
T \approx 10^{26} \left({M\over g}\right)^{-1}\ {\rm K}
\approx \left({M\over 10^{13} g}\right)^{-1} \ {\rm 
GeV}.
   \ee
This assumes that the hole has no charge or angular
momentum. This is a reasonable assumption since charge and 
angular
momentum will also be lost through quantum emission but on a 
shorter
timescale that the mass \cite{p77}. This means that it loses mass at a rate
   \be
      \dot{M} = -5\times 10^{25}(M/g)^{-2}f(M)~\mbox{g~s}^{-1}
   \ee
where the factor $f(M)$ depends on the number of particle 
species which are light enough to be emitted by a hole of mass 
$M$, so the 
lifetime is
   \be
      \tau(M) = 6 \times 10^{-27} f (M)^{-1}(M/g)^3~\mbox{s}.
   \ee
The factor $f$ is normalized to be 1 for holes larger than 
$10^{17}$~g
and such holes are only able to emit ``massless" particles like
photons, neutrinos and gravitons. Holes in the mass range
$10^{15}~\mbox{g} < M < 10^{17}~\mbox{g}$ are also able to emit
electrons, while those in the range $10^{14}~\mbox{g} < M <
10^{15}~\mbox{g}$ emit muons which subsequently decay into 
electrons
and neutrinos. The latter range includes, in particular, the critical
mass for which $\tau$ equals the age of the Universe. If the total density parameter is 1, this can be 
shown to $M_* = 4.4 \times 10^{14}h^{-0.3}$g where $h$ is the Hubble parameter
in units of 100 \cite{mc}.

Once $M$ falls below $10^{14}$g, a black hole can also begin to emit
hadrons. However, hadrons are composite particles made up of 
quarks
held together by gluons. For temperatures exceeding the QCD 
confinement scale of $\Lambda_{\rm QCD} = 250-300$~GeV, one 
would
therefore expect these fundamental particles to be emitted rather 
than
composite particles.  Only pions would be light enough to be 
emitted
below $\Lambda_{\rm QCD}$. Since there are 12 quark degrees of 
freedom per
flavour and 16 gluon degrees of freedom, one would also expect 
the
emission rate (i.e. the value of $f$) to increase dramatically once 
the
QCD temperature is reached.

The physics of quark and gluon emission from black holes is 
simplified
by a number of factors. Firstly, one can 
show that the separation between successively emitted particles is about 20 times 
their wavelength, which
means that short range interactions between them can be neglected. Secondly, the condition 
$T>\Lambda_{\rm QCD}$
implies that their separation is much less than $\Lambda_{\rm 
QCD}^{-1}
\approx 10^{-13}$cm (the characteristic strong interaction range) 
and
this means that the particles are also unaffected by strong
interactions. The implication of these three conditions is that one 
can
regard the black hole as emitting quark and gluon jets of the kind
produced in collider events.  The jets will decay into hadrons over 
a distance which is always much larger than $GM$, so gravitational 
effects can be neglected. The hadrons may then decay into astrophysically stable particles through weak and electomagnetic decays.

To find the final spectra of stable particles emitted from a black
hole, one must convolve the Hawking emission spectrum with the jet fragmentation function. This gives the instantaneous
emission spectrum shown in Fig.4 for a $T=1$~GeV black 
hole \cite{mw}. The direct emission just 
corresponds to
the small bumps on the right. All the particle spectra show a peak at
100~MeV due to pion decays; the electrons and neutrinos also 
have peaks at 1~MeV due to neutron decays.
In order to determine the present day background spectrum of
particles generated by PBH evaporations, one must first 
integrate over
the lifetime of each hole of mass $M$ and then over the PBH 
mass
spectrum \cite{mw}. In doing this, one must allow for the 
fact
that smaller holes will evaporate at an earlier cosmological epoch, 
so
the particles they generate will be redshifted in energy by the
present epoch. 

\begin{figure}
\centerline{
\includegraphics[width=10cm]{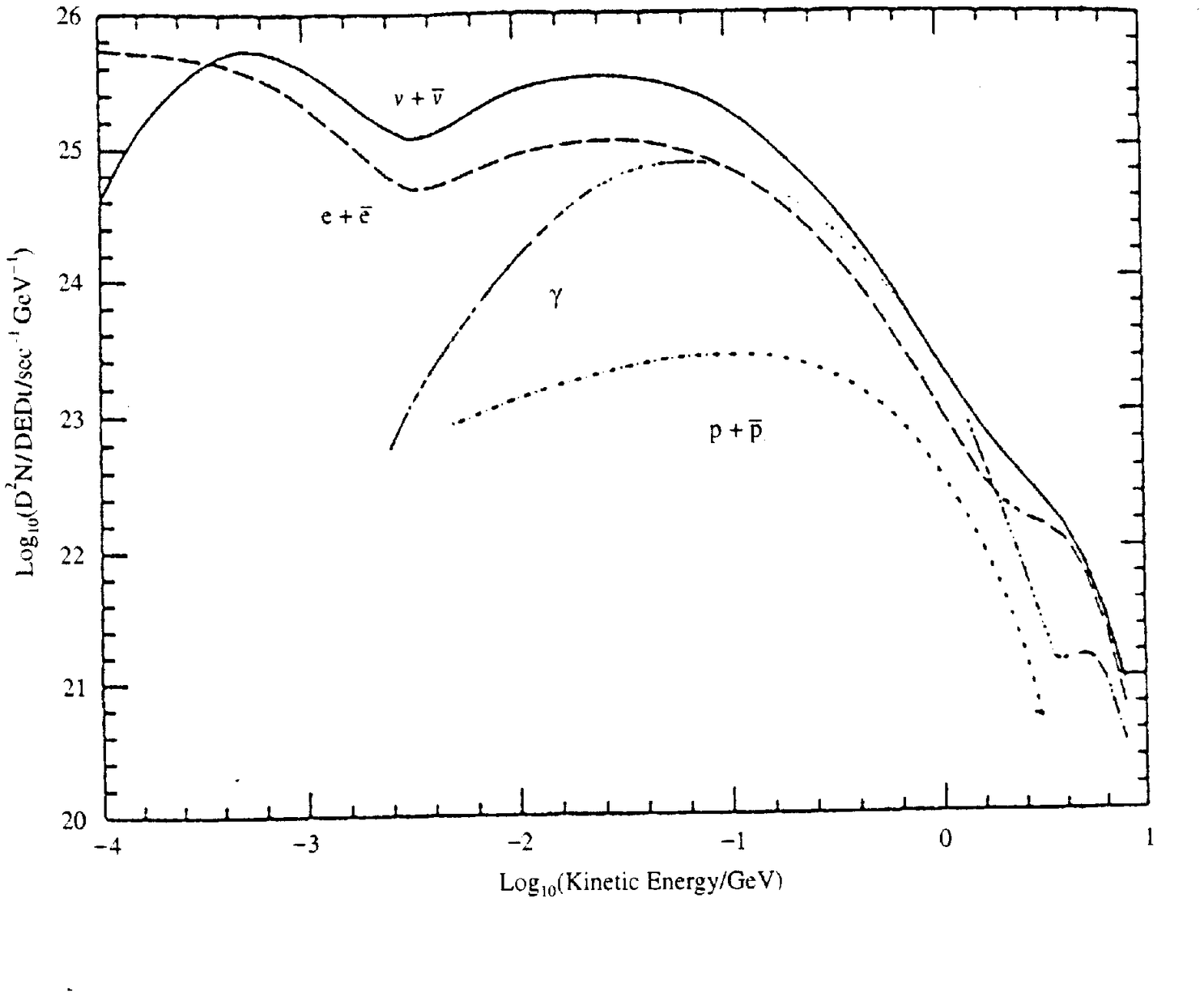}}
\caption{Instantaneous emission from a 1~GeV black hole}
\end{figure}

If the holes are uniformly distributed throughout the
Universe, the background spectra should have the form indicated 
in Fig.5. All the spectra have rather similar shapes: an $E^{-3}$
fall-off for $E > 100$~MeV due to the final phases of evaporation 
at the present epoch and an $E^{-1}$ tail for $E<100$~MeV due to the
fragmentation of jets produced at the present and earlier epochs. 
Note
that the $E^{-1}$ tail generally masks any effect associated with the mass spectrum of smaller PBHs which evaporated at earlier epochs \cite{c76}.

The situation is more complicated if the PBHs evaporating at 
the present epoch are clustered inside our own Galactic halo (as is 
most
likely). In this case, any charged particles emitted after the epoch
of galaxy formation (i.e. from PBHs only somewhat smaller than $M_*$) will have their flux enhanced relative to the
photon spectra by a factor $\xi$ which depends upon the halo
concentration factor and the time for which particles are trapped
inside the halo by the Galactic magnetic field.
This time is rather uncertain and also energy-dependent. At 100~MeV
one has $\xi \sim 10^3$ for electrons or positrons 
and $\xi \sim
10^4$ for protons and antiprotons. MacGibbon \& Carr \cite{mc} first used the observed cosmic ray spectra to constrain $\Omega_{\rm PBH}$ but their estimates have recently been updated.

 \begin{figure}
\centerline{
\includegraphics[width=10cm]{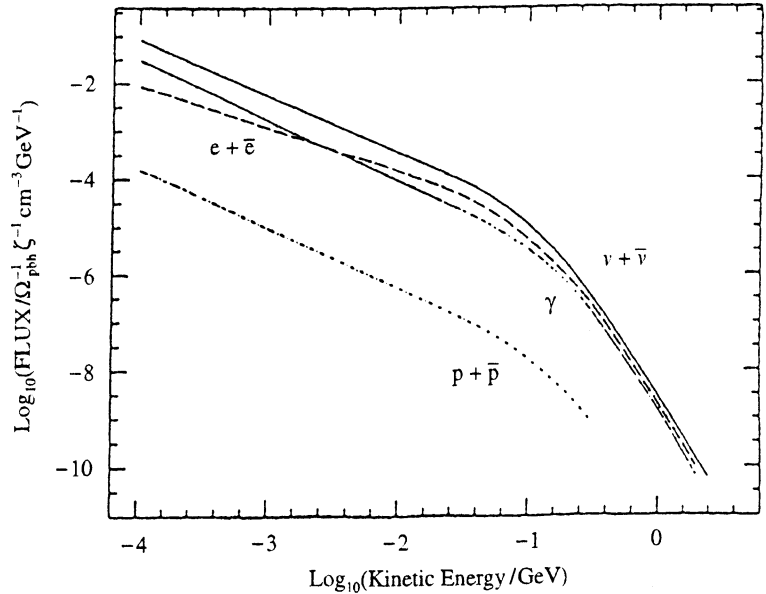}}
\caption{Spectrum of particles from uniformly distributed PBHs}
\end{figure}

\subsection{Gamma-rays}
Recent EGRET observations \cite{sre} give a $\gamma$-ray background of 
   \be
{dF_\gamma \over dE} = 
 7.3(\pm 0.7)\times 10^{-14}\left({E\over100MeV}\right)^{-2.10\pm 0.03}{\rm 
cm}^{-3} {\rm GeV}^{-1}
   \ee
between 30 MeV and 120 GeV. Carr \& MacGibbon \cite{cm} showed that this leads to an upper limit 
   \be
\Omega_{\rm PBH} \le (5.1 \pm1.3) \times 10^{-9} h^{-2}, 
   \ee
which is a refinement of the original Page-Hawking limit, but the form of the spectrum suggests that PBHs do not
provide the dominant contribution. 
If PBHs are clustered inside our own Galactic halo, then there should also be a Galactic $\gamma$-ray background and, since
this would be anisotropic, it  should be separable from the extragalactic background. The ratio of the anisotropic to
isotropic intensity depends on the Galactic longtitude and latitude, the ratio of the core radius to our Galactocentric radius, and the halo flattening. Wright claims that such a halo background has been detected \cite{w96}. His detailed fit to the EGRET data, subtracting various other known components, requires the PBH clustering factor to be $(2-12) \times 10^5 h^{-1}$, comparable to that expected.

\subsection{Antiprotons}
Since the ratio of antiprotons to protons in cosmic rays is less than
$10^{-4}$ over the energy range $100~\mbox{MeV} - 
10~\mbox{GeV}$,
whereas PBHs should produce them in equal numbers, PBHs could only
contribute appreciably to the antiprotons \cite{t82}. It is usually assumed 
that the observed
antiproton cosmic rays are secondary particles, produced by 
spallation
of the interstellar medium by primary cosmic rays. However, 
the spectrum of secondary antiprotons should show a steep
cut-off at kinetic energies below 2 GeV, whereas the spectrum of PBH antiprotons should increase with decreasing energy down to 0.2 GeV, so this provides
a distinct signature \cite{kir}. 

MacGibbon \& Carr originally calculated the PBH density required to explain the interstellar antiproton flux at 
1 GeV and found a value somewhat larger than the $\gamma$-ray limit \cite{mc}. More recent data on the
antiproton flux below 0.5 GeV comes from the BESS balloon experiment \cite{yosh} and Maki \etal \cite{mmo} have tried to fit this data in the  
PBH scenario. They model the Galaxy as a cylindrical diffusing halo of diameter 40 kpc
and thickness 4-8 kpc and then using Monte Carlo
simulations of cosmic ray propagation. A comparison with the data shows
no positive evidence for PBHs (i.e. there is no tendency for the antiproton fraction to tend to 0.5 at low energies) but they require the fraction of the 
local halo density in PBHs to be less than $3 \times10^{-8}$
and this is stronger than the $\gamma$-ray background limit.
A more recent attempt to fit the observed antiproton spectrum with PBH emission  comes from Barrau et al. \cite{bar} and is shown in Fig.6. A key test of the PBH hypothesis will arise during the 
solar minimum period because the flux of primary
antiprotons should be enhanced then, while that of the secondary
antiprotons should be little affected \cite{mos}.

\begin{figure}
\centerline{
\includegraphics[width=10cm]{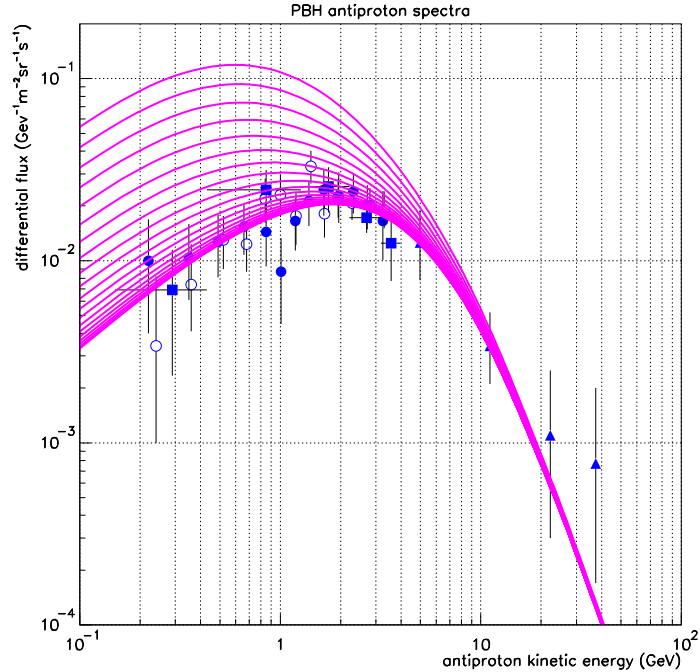}}
\caption{Comparison of PBH emission and antiproton data from
Barrau \etal}
\end{figure}

\subsection{PBH Explosions}
One of the most striking observational consequences of PBH evaporations would be their final explosive phase. However, 
in the standard particle physics picture, where the number of elementary particle species never exceeds around 100, the
likelihood of detecting such explosions is very low. Indeed, in 
this case, observations only place an upper limit
on the explosion rate of $ 5 \times10^8 {\rm pc}^{-3}{\rm y}^{-1}$ \cite{alex,sem}. This 
compares to Wright's $\gamma$-ray halo limit of 0.3 ${\rm pc}^{-3}{\rm y}^{-1}$ and the Maki \etal antiproton limit of 0.02 ${\rm pc}^{-3}{\rm y}^{-1}$.

However, the physics at the QCD phase transition is
still uncertain and the prospects of detecting explosions would
be improved in less conventional particle physics models. For example, in a Hagedorn-type picture, where the number of particle species exponentiates at the the quark-hadron temperature, 
the upper limit is reduced to 0.05 ${\rm pc}^{-3}{\rm y}^{-1}$ \cite{fic}.
Cline and colleagues have argued that one might
expect the formation of a QCD fireball at this temperature \cite{cho} and this might even explain some of the short period $\gamma$-ray bursts observed by BATSE \cite{csh}.
They claim to have found 42 candidates of this kind and the fact that their distribution matches the spiral arms suggests that they are Galactic. Although this proposal is speculative and has been disputed \cite{g01}, it has the attraction of making testable predictions (eg. the
hardness ratio should increase as the duration of the burst decreases). A rather different way of producing a $\gamma$-ray burst is to assume that the outgoing charged particles form a plasma due to turbulent magnetic 
field effects at sufficiently high temperatures \cite{bel}.

Some people have emphasized the possibility of detecting
very high energy cosmic rays from PBHs using air shower techniques \cite{css,hzmw,klw}. However, recently these 
efforts have been set back by the claim of Heckler \cite{hec1} that QED interactions could
produce an optically thick photosphere once the black hole temperature exceeds $T_{crit}=45$ GeV. In this case, the mean photon
energy is reduced to $m_e(T_{BH}/T_{crit})^{1/2}$, which is well below $T_{BH}$, so
the number of high energy photons is much reduced. He
has proposed that a similar effect may operate at even lower temperatures due to QCD effects \cite{hec2}. Several groups have examined the implications of this proposal for PBH emission \cite{cms,kap}. However, these 
arguments should not be regarded as definitive since MacGibbon \etal claim that QED and QCD interactions are never important \cite{mcp}.

\section{PBHs as a probe of quantum gravity}

In the standard Kaluza-Kelin picture, the extra dimensions are assumed to be compactified on the scale of the Planck length. This means that the influence of these extra dimensions only becomes important at an energy scale of $10^{19}$GeV and this is also presumbaly the scale on which quantum gravity effects become significant. In particular, such effects are only important for black hole evaporations once the black hole mass gets down to the Planck mass of $10^{-5}$g. Conceivably, this could result in black hole evaporation ceasing, so that one ends up with stable Planck-mass relics, and this leads to the sort of ``relics" constraints indicated in Fig.1, Fig.2 and Fig.3.  Various non-quantum-gravitational effects (such as higher order corrections to the gravitational Lagrangian or string effects) could also lead to stable relics \cite{cgl} but the relic mass is always close to the Planck mass.

In ``brane" versions of Kaluza-Klein theory, some of the extra dimensions can be much larger than the Planck length and this means that quantum gravity effects may become important at a much smaller energy scale.
If the internal space has $n$ dimensions and a compact volume $V_n$,  then Newton's constant $G_N$ is related to the higher dimensional gravitational constant $G_D$ and the value of the modified Planck mass $M_P$ is related to the usual 4-dimensional Planck mass $M_4$ by the order-of-magnitude equations:
\be
G_N \sim G_D/V_n, \quad M_P^{n+2} \sim M_4^2/V_n.
\ee
The same relationship applies if one has an infinite extra dimension but with a ``warped" geometry, provided one interprets $V_n$ as the ``warped volume". In the standard model, $V_n \sim 1/M_4^n$ and so $M_P \sim M_4$. However, with large extra dimensions, one has $V_n >> 1/M_4^n$ and so $M_P << M_4$. In particular, this might permit quantum gravitational effects to arise at the experimentally observable TeV scale.

If this were true, it would have profound implications for black hole formation and evaporation since black holes could be generated in accelerator experiments, such as the Large Hadron Collider (LHC). Two partons with centre-of-mass energy $\sqrt{s}$ will form a black hole if they come within a distance corresponding to the Schwarzschild radius $r_S$ for a black hole whose mass $M_{BH}$ is equivalent to that energy \cite{dl,gt,rt}. Thus the cross-section for black hole production is
\be
\sigma_{BH} \approx \pi r_S^2 \Theta(\sqrt{s}-M_{BH}^{min})
\ee
where $M_{BH}^{min}$ is the mass below which the semi-classical approximation fails. Here the Schwarzschild radius itself depends upon the number of internal dimensions:
\be
r_S \approx {1\over M_P}\left({M_{BH}\over M_P}\right)^{1/(1+n)},
\ee
so that $\sigma_{BH} \propto s^{1/(n+1)}$. This means that the cross-section for black hole production in scattering experiments goes well above the cross-section for the standard model above a certain energy scale and in a way which depends on the number of extra dimensions.

The evaporation of the black holes produced in this way will produce a characteristic signature \cite{dl,gt,rt} because the temperature and lifetime of the black holes depend on the number of internal dimensions:
\be
T_{BH} \approx {n+1\over r_S}, \quad \tau_{BH} \approx {1\over M_P}\left({M_{BH}\over M_P}\right)^{(n+3)/(n+1)}.
\ee
Thus the temperature is decreased relative to the standard 4-dimensional case and the lifetime is increased. The important qualitative effect is that a large fraction of the beam energy is converted into transverse energy, leading to large-multiplicity events with many more hard jets and leptons than would otherwise be expected. In principle, the formation and evaporation of black holes might be observed by LHC by the end of the decade and this might also allow one to experimentally probe the number of extra dimensions. On the other hand, this would also mean that scattering processes above the Planck scale could not be probed directly because they would be hidden behind a black hole event horizon.  

Similar effects could be evident in the interaction between high energy cosmic rays and atmospheric nucleons. Nearly horizontal cosmic ray neutrinos would lead to the production of black holes, whose decays could generate deeply penetrating showers  with an electromagnetic component substantially larger than that expected with conventional neutrino interactions. Several authors have studied this in the context of the Pierre Auger experiment, with event rates in excess of one per year being predicted \cite{ag,fs,rt}. Indeed there is a small window of opportunity in which Auger might detect such events before LMC.

It should be stressed that the black holes produced in these processes should not themselves be described as ``primordial" since they do not form in the early Universe. On the other hand, it is clear that the theories which predict such processes will also have profound implications for the formation and evaporation of those black holes which {\it do} form then. This is because, at sufficiently early times, the effects of the extra dimensions must be cosmologically important. However, these effects are not  yet fully understood.

\section{Conclusions}

We have seen that PBHs could provide a unique probe of the early Universe, gravitational collapse, high energy physics and quantum gravity. In the ``early Universe" context, particularly useful constraints can be placed on inflationary scenarios and on models in which the value of the gravitational ``constant" G varies with cosmological epoch. In the ``gravitational collapse" context, the existence of PBHs could provide a unique test of the sort of critical phenomena discovered in recent numerical calculations. In the ``high energy physics" context, information may come from observing cosmic rays from evaporating PBHs since the constraints on the number of evaporating PBHs imposed by gamma-ray background observations do not exclude their making a significant contribution to the Galactic flux of electrons, positrons and antiprotons. Evaporating PBHs may also be detectable in their final explosive phase as gamma-ray bursts if suitable physics is invoked at the QCD phase transition. In the ``quantum gravity" context, the formation and evaporation of small black holes could lead to observable signatures in cosmic ray events and accelerator experiments, providing there are extra dimensions and providing the quantum gravity scale is around a TeV.

\end{document}